\documentclass[runningheads]{svmult}

\usepackage{makeidx}   
\usepackage{graphicx}  
\usepackage{subeqnar}  
\usepackage{multicol}  
\usepackage{physprbb}  
\makeindex             

\begin{document}
\title*{Why Do AGN Lighthouses Switch Off?}
\titlerunning{Why Do AGN Lighthouses Switch Off?}
%
\author{Ramesh Narayan}
\authorrunning{Ramesh Narayan}
%
%
\institute{Harvard-Smithsonian Center for Astrophysics,
           60 Garden Street, Cambridge, MA 02138, U.S.A.}

\maketitle              

\begin{abstract}
Nearby galactic nuclei are observed to be very much dimmer than active
galactic nuclei in distant galaxies.  The Chandra X-ray Observatory
has provided a definitive explanation for why this is so.  With its
excellent angular resolution, Chandra has imaged hot X-ray-emitting
gas close to the gravitational capture radius of a handful of
supermassive black holes, including Sgr A$^*$ in the nucleus of our
own Galaxy.  These observations provide direct and reliable estimates
of the Bondi mass accretion rate $\dot M_{Bondi}$ in these nuclei.  It
is found that $\dot M_{Bondi}$ is significantly below the Eddington
mass accretion rate, but this alone does not explain the dimness of
the accretion flows.  In all the systems observed so far, the
accretion luminosity $L_{acc}\ll 0.1\dot M_{Bondi}c^2$, which means
that the accretion must occur via a radiatively inefficient mode.
This conclusion, which was strongly suspected for many years, is now
inescapable.  Furthermore, if the accretion in these nuclei occurs via
either a Bondi flow or an advection-dominated accretion flow, the
accreting plasma must be two-temperature at small radii, and the
central mass must have an event horizon.  Convection, winds and jets
may play a role, but observations do not yet permit definite
conclusions.

\end{abstract}

\section{Introduction}
At high redshift, many galactic nuclei are extremely bright, with
luminosities in excess of $10^{45}~{\rm erg\,s^{-1}}$.  These active
galactic nuclei (AGN) are believed to be powered by accretion of gas
onto supermassive black holes (SMBHs) at nearly the Eddington rate
\cite{K99}.  The accretion very likely occurs via a Shakura-Sunyaev
thin disk \cite{SS73},\cite{NT73},\cite{P81},\cite{FKR92}.  The Big
Blue Bump, which is present in the spectra of all bright AGN, is
identified with blackbody emission from the optically thick disk
\cite{KB99}, while the X-ray emission (roughly 10\% of the luminosity)
is thought to be produced by an optically thin corona above the disk
\cite{HM91},\cite{HM93}.  In addition, most AGN have substantial
infrared emission, usually the result of dust reprocessing at a
relatively large distance from the SMBH.  Some AGN also have
significant radio emission from relativistic jets.

Most galactic nuclei at low redshift are very different.  These nearby
nuclei are much less active --- sometimes not active at all.  The
nuclear source in our own Galaxy, Sagittarius A$^*$, is a particularly
good example of a dim galactic nucleus.  Studies of AGN demographics
suggest that a significant fraction of (perhaps all?)  SMBHs must have
gone through an AGN ``lighthouse'' phase at some early stage in their
lives.  Why and how did these early lighthouses switch off to become
the dormant nuclei we see today?

The reduced activity in nearby galactic nuclei is certainly not
because of a lack of SMBHs.  Recent observations have provided ample
evidence that virtually every galaxy in our local neighborhood has a
SMBH \cite{Retal98}.  The lack of activity must therefore be the
result of reduced gas supply.  While this is certainly part of the
answer, we shall see that it is not the whole story.  A second,
equally important, reason is that the very mode of accretion is
different in dim nuclei: the accretion occurs via a radiatively
inefficient mode, such that even what little gas reaches the SMBH
produces much less radiated energy per unit accreted mass than in
bright AGN.  We review below the evidence for this conclusion.  Our
primary emphasis is on the nearest and best-studied dim SMBH, Sgr
A$^*$, but we also discuss briefly other dim nuclei.  The focus is on
the observations and on what we can or cannot tell with confidence
from the presently available data.

\section{Observations of Sgr A$^*$}

\subsection{Spectrum}
\begin{figure}[b]
\begin{center}
\includegraphics[width=.6\textwidth]{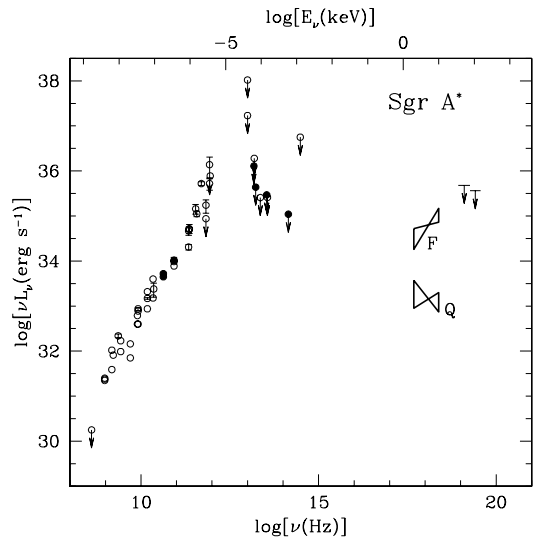}
\end{center}
\caption[]{Spectral data on Sgr A$^*$.  Shown is $\nu L_\nu$, the
luminosity per logarithmic interval of frequency $\nu$, versus $\nu$
(lower scale) or photon energy $E_\nu=h\nu$ (upper scale).}
\label{sgrdata}
\end{figure}
Figure 1 summarizes the available data on the spectrum of Sgr A$^*$
\cite{MDZ96},\cite{Netal98},\cite{MF01},\cite{Betal01a},\cite{Betal01b}.
Until recently, Sgr A$^*$ had been firmly detected only in the radio
and millimeter bands, where the source has a hard rising spectrum.
There are useful upper limits in the submillimeter and infrared bands
which indicate that the emission peaks in the submillimeter/far
infrared region of the spectrum, with a luminosity of about
$10^{36}~{\rm erg\,s^{-1}}$.  Sgr A$^*$ may have been detected during
one observation at 2.2 $\mu$m \cite{Getal97}, with a luminosity of
$10^{35}~{\rm erg\,s^{-1}}$, but the detection was not confirmed
in other observations.  If the source is variable, the lone
detection may represent its brightest state.  The data point is shown
as an upper limit in Fig. 1.  In the optical, UV and soft X-ray bands,
the source is strongly extincted (visual extinction $A_V\approx 30$
magnitudes \cite{MS96}) and there are no direct constraints on the
spectrum.  Nevertheless, it is generally agreed that the emission in
these hidden bands is unlikely to exceed that in the submillimeter
peak.

Over the years, Sgr A$^*$ has been observed many times in X-rays
\cite{Wetal81},\cite{PT94},\cite{Getal94},\cite{Ketal96}.  However,
because the source is very weak and the field is highly crowded, there
was no unambiguous detection until the recent observations of Baganoff
et al. \cite{Betal01a},\cite{Betal01b} with the Chandra X-ray
Observatory.  The exquisite angular resolution of Chandra allowed Sgr
A$^*$ to be isolated within the crowded field and its flux and
spectrum to be measured.  Out of a total of 76 ks of observations,
extending over two epochs, the source was in a quiescent state (marked
Q in Fig. 1) for nearly 70 ks, with a flux of $2.2\times10^{33}~{\rm
erg\,s^{-1}}$ and a relatively soft spectrum (photon index
$\Gamma\sim1.6-2.8$ \cite{Betal01b}).  For a brief period of several
ks during the second epoch of observation, the source went into a
flare state (F in Fig. 1), during which the flux went up by a factor
of a few tens and the spectrum became quite hard
($\Gamma\sim0.7-1.8$).  Integrated over time, the emission in the
flare was small compared to the quiescent emission.  We therefore take
the quiescent data as representative of the average properties of the
source.  We discuss the flare briefly in \S5.3.

The mass of the SMBH at the center of our Galaxy has been measured to
be $2.6\times10^6 M_\odot$
\cite{EG97},\cite{Ghetal98},\cite{Getal00},\cite{Ghetal00}, which
means that the Eddington luminosity of Sgr A$^*$ is
$L_{Edd}=3\times10^{44}~{\rm erg\,s^{-1}}$.  On the scale of
$L_{Edd}$, the peak emission in the submillimeter is
$L_{submm}\sim10^{-8.5}L_{Edd}$.  The luminosity in the infrared is
even less, $L_{IR}<10^{-9.5}L_{Edd}$, and the quiescent X-ray
luminosity is a pitiful $L_X\sim10^{-11}L_{Edd}$.  Thus, Sgr A$^*$ is
an extremely dim galactic nucleus.  Indeed, it is the dimmest nucleus
for which we have useful data, which explains why this source plays
such a central role in all discussions of dim galactic nuclei.

\subsection{Bondi Accretion Rate}
In a famous paper, Bondi \cite{B52} discussed the problem of spherical
hydrodynamical accretion onto a black hole (BH) of mass $M$ immersed
in a uniform medium of density $\rho_0$ and sound speed $c_{s,0}$.  He
showed that the sphere of influence of the BH extends out to the
gravitational capture radius $R_c=GM/c_{s,0}^2$.  Gas external to
$R_c$ is only mildly perturbed by the BH, whereas any gas that falls
within $R_c$ is gravitationally captured by the BH and free-falls down
to the center.  The radial velocity in the free-fall zone is
\begin{equation}
v_R \sim (2GM/R)^{1/2}\sim c_{s,0}(R/R_c)^{-1/2},
\end{equation}
and so the mass accretion rate onto the BH is approximately (see
Bondi's original article or \cite{ST83} for exact results)
\begin{equation}
\dot M_{Bondi}\sim 4\pi R_c^2\rho_0c_{s,0}.
\end{equation}
Bondi's model applies strictly to a uniform infinitely extended
medium.  However, even if the external medium is not truly uniform,
the above formula for $\dot M_{Bondi}$ is still valid, provided we use
for $\rho_0$ the density near the capture radius $R_c$.

Until recently, the Bondi accretion rate in Sgr A$^*$ could be
estimated only indirectly, by estimating what fraction of the winds
ejected by surrounding stars is captured by the SMBH (see
\cite{CM97},\cite{QNR99}).  The estimate had large uncertainties.
Chandra has changed the situation dramatically by directly imaging hot
X-ray-emitting thermal gas in the vicinity of the capture radius of
the SMBH \cite{Betal01a}.  The observations show that there is
extended $\sim1$ keV gas with a number density $n_0\sim 30~{\rm
cm^{-3}}$ over an area in the sky of about 10 arcsec around Sgr A$^*$.
In addition, there is $\sim2$ keV gas with $n_0\sim100~{\rm cm^{-3}}$
spread over about an arcsec around the source (and clearly resolved by
Chandra \cite{Betal01a}).  The capture radius of Sgr A$^*$ for keV gas
is about 1 arcsec (for a BH mass of $2.6\times10^6M_\odot$ and a
distance of 8.5 kpc), so the 2 keV emission presumably comes from gas
that has just been captured by the BH and been mildly heated by
compression.  Thus, Chandra provides a direct measurement of the
properties of the gas ($n_0$ and $c_{s,0}$) right at the capture
radius.  Using this, one can obtain a reliable estimate of the Bondi
accretion rate \cite{Betal01a}:
\begin{equation}
\dot M_{Bondi} \sim (0.3-1)\times10^{-5}M_\odot{\rm
yr^{-1}}\sim 3\times10^{20}~{\rm g\,s^{-1}}\sim 10^{-4}\dot
M_{Edd},
\end{equation}
where we have formally assumed a radiative efficiency of 10\% in the
definition of the Eddington accretion rate, i.e. $\dot M_{Edd}\equiv
L_{Edd}/0.1c^2$.

\section{There is No Thin Disk in Sgr A$^*$}
\begin{figure}[b]
\begin{center}
\includegraphics[width=.6\textwidth]{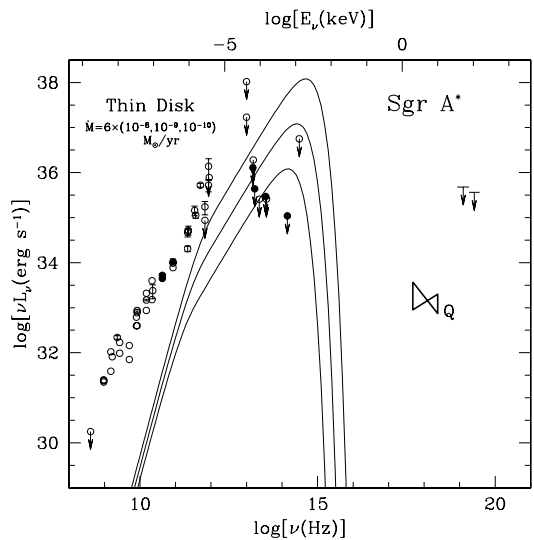}
\end{center}
\caption[]{Spectra corresponding to thin disk models of Sgr A$^*$ with
three choices of $\dot M$.  All three models shown have $\dot M\ll
\dot M_{Bondi}$.  Yet, all three models are too bright to fit the
data.}
\label{sgrthin}
\end{figure}
Could Sgr A$^*$ have a Shakura-Sunyaev thin disk?  In the most
straightforward version of this model, mass would flow onto the disk
at the Bondi rate $\dot M_{Bondi}$ estimated in (3) and would flow
steadily through the disk onto the SMBH.  The model would predict a
luminosity $L_{disk}\sim 0.1\dot M_{Bondi}c^2 \sim 10^{40.5}~{\rm
erg\,s^{-1}}$, with the bulk of the emission appearing in the near
infrared and optical bands.  Figure 2 shows model spectra
corresponding to a thin disk with three choices of $\dot M$:
$(10^{-6}, 10^{-7}, 10^{-8})\times\dot M_{Edd}$.  Even models with
such low accretion rates, which are far lower than $\dot M_{Bondi}$,
are ruled out by the infrared limits.

One way of trying to save the thin disk model is to assume that gas
flows in at the Bondi rate at $R_c$, but then condenses onto a cold
``dead'' disk where it sits without accreting onto the BH.  This would
require the disk to be in a very quiet state with an extremely low
viscosity.  This is not unreasonable --- for instance, cataclysmic
variables and soft X-ray transient binaries have quiescent states in
which the gas accretes onto the central mass at a much lower rate than
the rate at which gas is fed on the outside by mass transfer from the
companion star.  In the case of Sgr A$^*$, however, such a model runs
into difficulties because the inflowing gas would produce a fair
amount of luminosity in the infrared, and possibly also X-rays, as it
crashes onto the thin disk and loses its thermal and kinetic energy
\cite{FM97}.  For example, cataclysmic variables and X-ray transients
in quiescence have optical and UV emission from the ``hot spot'' where
the incoming gas stream hits the disk, even though the disk itself may
be hardly accreting.  The lack of any evidence for thermalization
radiation from the inflowing gas rules out the thin disk model quite
strongly in Sgr A$^*$.  In the opinion of this author, any variant of
the model that succeeds in getting round this constraint is likely to
be very contrived.  We will therefore take it as given that there is
no thin disk in Sgr A$^*$.

As an aside, we note that the above argument does not apply if the
accretion occurs via an advection-dominated accretion flow (see
\S\S5.1,5.2) instead of a thin disk (despite claims to the contrary
\cite{MF01}).  The reason is that in the case of an advective flow,
there is no free-falling Bondi-like zone.  At the capture radius the
gas directly makes a transition to a subsonic rotating accretion flow,
and there is no supersonic zone or shock.  (In addition, even if there
were a shock it is not clear that the hot gas would radiate very much,
as it is advection-dominated.)

Returning to the discussion on thin disk models, the essence of the
argument against the presence of a thin disk in Sgr A$^*$ is the fact
that the accretion luminosity $L_{acc}$ is very low,
\begin{equation}
L_{acc}\ll 0.1\dot M_{Bondi}c^2,
\end{equation}
whereas a thin disk, being radiatively efficient, will normally have
$L_{acc}\sim 0.1\dot M_{Bondi}c^2$.  If Sgr A$^*$ does not have a thin
disk, then what kind of a flow does it have?  Whatever the flow is,
the data tell us that it has to be radiatively highly inefficient.
Radiatively inefficient gas will be hot, because the energy has
nowhere to go except into thermal energy.  The gas will also be
quasi-spherical rather than thin.  We now turn to a consideration of
models with these properties.

\section{Bondi Model}

\subsection{Spherical Accretion}
The most famous example of a radiatively inefficient accretion flow is
Bondi spherical accretion \cite{B52}.  The radial velocity of the
accreting gas and the mass accretion rate in this model are given by
(1) and (2) above.  From these, the density profile is easily
obtained: $\rho\sim\rho_0(R/R_c)^{-3/2}$, where $\rho_0$ is the
density of the external medium.

Bondi originally assumed that the accreting gas is neither heated nor
cooled.  This assumption is not valid if the gas is magnetized.  As
Shvartsman \cite{S71} first argued, any magnetic field frozen in the
gas is amplified as $B\propto R^{-2}$, and so the magnetic energy
density grows as $R^{-4}$.  The gas energy density, however, varies
only as $R^{-5/2}$, which means that even if the gas starts off with a
sub-equipartition strength $B$, the field quickly grows to
equipartition strength at some radius.  Inside this radius, the field
will reconnect so as to maintain rough equipartition.  The
reconnection will heat the gas \cite{M75}, and as a result, the
thermal energy will come roughly into equipartition with the magnetic
energy.  Since there is negligible cooling, the sum of the thermal,
magnetic and kinetic energies should equal the potential energy of the
gas.  It is then easy to show that the temperature of the gas will
vary roughly as
\begin{equation}
T\sim 10^{12}(R/R_S)^{-1}~{\rm K}, \qquad c_s\sim c\, (R/R_S)^{-1/2},
\end{equation}
where $R_S=7.7\times10^{11}(M/2.6\times10^6M_\odot)~{\rm cm}$ is the
Schwarzschild radius of the accreting BH.

The heating of the gas leads to convection; we discuss this topic in
\S7.1.

\subsection{Bondi Models of Sgr A$^*$}
\begin{figure}[b]
\begin{center}
\includegraphics[width=.6\textwidth]{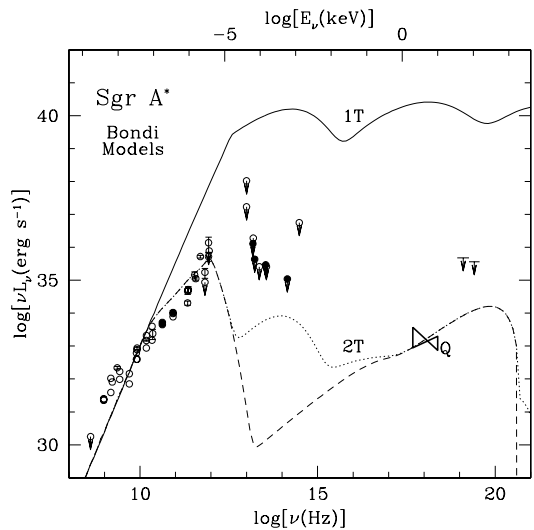}
\end{center}
\caption[]{Spectra of Sgr A$^*$ corresponding to Bondi spherical
accretion models (see the text for details)}
\label{sgrbondi}
\end{figure}
Given the above profiles of density and temperature, it is
straightforward to calculate the spectrum of a Bondi accretion flow.
The primary emission mechanisms are synchrotron and bremsstrahlung,
plus Comptonization of each.  The solid line in Fig. 3 shows the
predicted spectrum of Sgr A$^*$ if accretion occurs via a Bondi flow.
We have normalized the density of the model such that, at $R=R_c$, the
density is equal to $\rho_0$ as measured by Chandra.  The model
clearly predicts far too much flux all across the spectrum.  The
reason is the very high temperature of the electrons, especially close
to the BH (see eq 5).  The hot electrons radiate synchro-Compton
emission very efficiently.

A Bondi model for Sgr A$^*$ was proposed by Melia
\cite{M92},\cite{M94} many years ago and pursued in detail by his
group.  This work was the first to propose that a hot accretion flow
emitting synchrotron and bremsstrahlung radiation could explain the
strange properties of Sgr A$^*$.

The published models of Melia fit the observed spectrum quite well,
which is rather surprising considering that we find the Bondi model to
overpredict the luminosity by many orders of magnitude (solid line in
Fig. 3).  More surprising still, the Melia models used an accretion
rate 100 times larger than the one used here in Fig. 3, and yet agreed
with the observations.  One reason for the large discrepancy is an
error (at least in the early work) in the computation of the
synchrotron spectrum (see \cite{MNY96}).  Another reason could be that
Compton-scattering was not included in some of the models.  It is also
possible that the models did not correctly include heating due to
magnetic dissipation.  Strong heating is generic to any spherical flow
that falls over many decades of radius (\S4.1).  The heating can be
avoided if the magnetic field is extraordinarily weak in the external
medium so that the field does not grow to equipartition strength even
at the BH horizon, but this is very unlikely.  Alternatively, heating
would be weak if the field reconnects while it is still well below
equipartition, as proposed in some models \cite{KM99}.  But this
requires the reconnection to proceed at much faster than the Alfven
speed, which seems unlikely (e.g., \cite{LV99a},\cite{LV99b}).

The electrons in a Bondi model of Sgr A$^*$ are too luminous because
they are too hot.  This suggests a possible fix: make the accreting
plasma two-temperature, with the electrons much cooler than the
protons.  Two-temperature models of spherical flows were developed in
detail by Shapiro and co-workers for accretion flows onto neutron
stars \cite{SS75} and BHs \cite{SLE76}.  If one assumes that ions and
electrons couple only via Coulomb collisions --- a not unreasonable
assumption --- then at the low density of the accretion flow in Sgr
A$^*$, the plasma would automatically develop a two-temperature
structure.

To illustrate how strong an effect the electron temperature has on the
spectrum, we consider an artificial problem in which we modify the
electron temperature profile in the Bondi flow such that it is capped
at $10^{10}$ K.  The model predicts the spectrum shown by the dotted
line in Fig. 3 (the dashed line is the contribution due to
bremsstrahlung and synchrotron alone, without Comptonization).  This
model is clearly in much better agreement with the observations than
the one-temperature model (solid line).  We thus conclude that any
successful model of Sgr A$^*$ is likely to involve a two-temperature
plasma.

Before closing this section, we note that the spherical accretion
model, while good for developing insight into the properties of hot
flows, is unlikely to describe real accretion flows since it ignores
the angular momentum of the accreting gas.  In almost any accretion
flow, the gas is likely to possess sufficient angular momentum that it
would hit the centrifugal barrier at some radius $R_{cb}$ before it
can fall into the BH.  Inside $R_{cb}$, accretion is possible only if
there is some agency to transfer angular momentum outward; we will
call this agency ``viscosity'' though it is probably magnetic stresses
\cite{BH98}.  Some authors have proposed models in which they invoke a
Bondi flow from the capture radius $R_c$ down to $R_{cb}$, and they
then introduce a thin disk or some other kind of rotating solution
from $R_{cb}$ down to the marginally stable orbit (e.g.,
\cite{M94},\cite{MLC00}).  The problem with any such model is that
there is nowhere for the angular momentum to go.  A viscous rotating
flow can accrete only if it can get rid of angular momentum to the
outside, but angular momentum cannot be transported across supersonic
zones such as a Bondi flow \cite{P77},\cite{PN92}.  Therefore, if the
supersonically infalling gas in a Bondi-like flow is arrested by
centrifugal forces, then whatever rotating structure forms at the
center must extend out at least to the capture radius in order to be
able to transfer angular momentum to the outside.  Many models of Sgr
A$^*$ in the literature fail to satisfy this simple consistency
condition.  (The ADAF model discussed next does satisfy the
condition.)

\section{Advection-Dominated Accretion Flow (ADAF) Model}

\subsection{The ADAF Model}
The thin accretion disk is a radiatively efficient model in which all
the energy released through viscous dissipation is radiated away.  One
could describe it as a cooling-dominated accretion flow.  An
``advection-dominated accretion flow'' (ADAF) is one in which most of
the heat energy released by viscosity and compression is retained in
the gas and advected to the center, and only a very small fraction of
the energy is radiated.

Technically, any radiatively inefficient accretion flow, including the
Bondi flow, is advection-dominated.  However, the term ADAF is
conventionally applied only to a particular class of quasi-spherical
models that include rotation, viscosity and a two-temperature plasma
\cite{I77},\cite{Retal82},\cite{NY94},\cite{NY95a},\cite{NY95b},
\cite{Aetal95}\cite{Cetal95}.  We will follow this convention here.

The reason ADAFs are advection-dominated is that the accreting gas has
a low density (because of the low mass accretion rate) and the thermal
structure of the plasma is two-temperature.  Because of the low
density, very little of the heat energy in the ions gets transferred
to the electrons through Coulomb collisions \cite{SLE76},\cite{NY95b},
\cite{T01}.  Since the ions hardly radiate at all, they retain their
thermal energy and advect essentially all of it to the center.  The
ion temperature thus scales essentially as in (5) and becomes of order
$10^{12}$ K near the BH.  The electrons do radiate some of their
energy.  They also have a different equation of state once they become
relativistic (at small radii).  For both reasons, they are cooler than
the ions, and do not get much hotter than ${\rm few}\times10^{10}$ K
near the BH.  The lower temperature makes the electrons less efficient
radiators than in the one-temperature Bondi model shown in Fig. 3.
Indeed, at sufficiently low $\dot M$ (as in Sgr A$^*$), the electrons
become radiatively quite inefficient, though not as inefficient as the
ions, and advect a large part of their energy.  In this regime, the
overall radiative efficiency of the accretion flow can be very low.

Most ADAF models in the literature assume that the ions and electrons
have thermal energy distributions, which is reasonable under many
conditions \cite{MQ97}.  A few papers have considered the effects of
nonthermal particles, e.g., \cite{MNK97},\cite{M99},\cite{OPN00}.

Because the gas in an ADAF has angular momentum, accretion is driven
by viscosity (not just gravity as in the Bondi flow).  The radial
velocity of the gas then scales roughly as $v_R\sim\alpha(GM/R)^{1/2}$
\cite{NY94}, where $\alpha$ is the usual dimensionless viscosity
parameter \cite{SS73},\cite{FKR92}.  The velocity is thus lower than
in a Bondi flow by a factor $\sim\alpha$, whose typical value is
$\sim0.1$.  The lower radial velocity means that, for a given ambient
density in the external medium, $\dot M_{ADAF}$ in an ADAF is lower
than $\dot M_{Bondi}$ in a Bondi flow by a factor $\sim\alpha\sim0.1$.

The ADAF solution is allowed only for relatively low values of $\dot
M$, less than a few per cent of the Eddington rate for $\alpha\sim0.1$
\cite{Retal82},\cite{NY95b},\cite{EMN97} (but see \cite{Yetal00}).
This makes the solution a natural choice for modeling low-luminosity
accretion flows.  Unlike other hot accretion flow solutions
\cite{SLE76},\cite{P78}, the ADAF does not suffer from any serious
thermal or viscous instability
\cite{KAC96},\cite{Ketal97},\cite{WL96},\cite{W97}.

The reader is referred to the following reviews for more details of
the ADAF model and its application to dim accretion sources:
\cite{NMQ98},\cite{Ketal98},\cite{Q01},\cite{NGM01}.

\subsection{ADAF Models of Sgr A$^*$}
ADAF models of Sgr A$^*$ have been described in a number of papers in
the literature
\cite{NYM95},\cite{MMK97},\cite{Netal98},\cite{M98},\cite{M99},\cite{OPN00},\cite{M00}.
These studies have shown that it is possible to explain both the very
low luminosity of Sgr A$^*$ and its spectrum without invoking an
unreasonably low mass accretion rate.  The above studies were done
before the recent Chandra observations, when there was only a rough
estimate of the mass accretion rate.  Here we present some new results
using the Chandra measurement of the density and temperature of the
external medium.

\begin{figure}[b]
\begin{center}
\includegraphics[width=.6\textwidth]{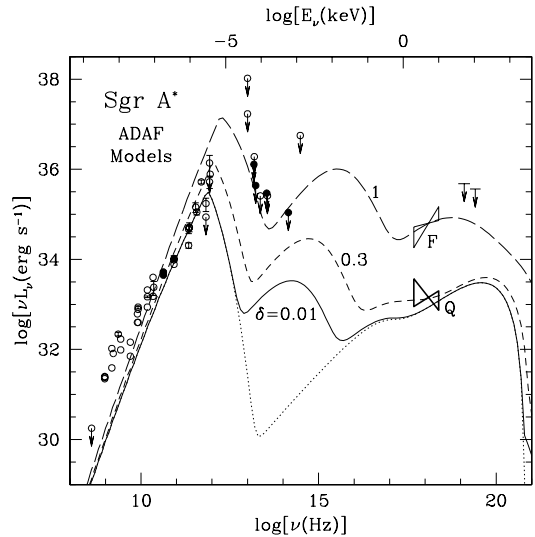}
\end{center}
\caption[]{ADAF models of Sgr A$^*$ (see the text for details)}
\label{sgradaf}
\end{figure}
Figure 4 shows spectra corresponding to three ADAF models of Sgr
A$^*$.  The models assume $\alpha=0.1$ (the results are not sensitive
to the precise value) and take the magnetic pressure in the gas to be
a tenth of the total pressure.  The density profile is matched to the
ambient density of the external medium at the capture radius; more
precisely, the models have been adjusted so that the predicted
bremsstrahlung emission agrees with the spatially resolved X-ray flux
measured by Chandra.  Because we have used a boundary condition on the
density, these ADAF models have a lower mass accretion rate, $\dot
M_{ADAF}\sim 10^{-6}M_\odot{\rm yr^{-1}}$, than the Bondi models
described in \S4.2.  This is to be expected since the radial velocity
in an ADAF is lower by a factor $\sim\alpha$ than in a Bondi flow.
Note that, because we allow the ADAF to extend all the way out to the
capture radius, there is no Bondi-like segment in the accretion flow.
The external gas directly makes a transition to the ADAF.  The flow is
subsonic all the way down to a sonic radius close to the BH, and there
is viscous coupling between the accretion flow and the external
medium.  This allows angular momentum to be expelled from the system
to the outside.

The three models shown in Fig. 4 differ in the value of a parameter
$\delta$, which measures the fraction of the viscous energy that goes
into the electrons.  Early ADAF models assumed that nearly all the
viscous heat goes into the ions and that $\delta$ is very small, say
$\sim0.01$.  The solid curve in Fig. 4 shows the spectrum for this
choice of $\delta$.  The spectrum was calculated by the methods
described in \cite{Netal98}.  We see that the calculated spectrum is
generally consistent with the data, though the X-ray spectrum may be a
little too hard \cite{Betal01a}.

Is it reasonable to assume that electrons receive such a small
fraction of the viscous heat energy \cite{Ph81}?  Several authors have
investigated this question
\cite{BL97},\cite{Q98},\cite{G98},\cite{QG99},\cite{B99}.  In brief,
while it appears that negligible electron heating is possible under
some conditions, under other conditions electrons are expected to
receive a good fraction (few tenths) of the energy.  The short-dashed
curve in Fig. 4 shows the predicted spectrum for a model with
$\delta=0.3$.  It is reassuring that this model, which uses a
``reasonable'' value of $\delta$, is again in pretty good agreement
with the data.  In fact, the model predicts a somewhat softer X-ray
spectrum than the $\delta=0.01$ model (because part of the emission is
now due to Comptonization), which agrees better with the Chandra data.

The dotted line in Fig. 4 corresponds to the contributions from
synchrotron and bremsstrahlung emission alone for the $\delta=0.01$
model.  The synchrotron peak on the left nicely fits the radio/mm
data, and cuts off where it should in order to satisfy the infrared
upper limits.  This emission comes from fairly small radii in the flow
(few to few tens of $R_S$).  The bremsstrahlung peak on the right
explains the quiescent X-ray emission in Sgr A$^*$ and is almost
entirely from large radii $\sim R_c\sim10^5R_S$, which corresponds to
an angular scale of order an arcsec.  The rest of the emission in the
spectrum is from Compton scattering, which fills in the region between
the synchrotron and bremsstrahlung peaks.  The Compton emission comes
primarily from radii inside $100R_S$.  It is very reassuring that the
model does not produce too much Compton emission in X-rays, since the
Compton emission would be spatially unresolved, whereas Chandra has
clearly shown that at least 50\% of the observed X-ray emission in
quiescence is resolved.  Overall, we conclude that an ADAF model with
$\delta<0.3$ is consistent with all the presently available data on
Sgr A$^*$.  Note that, since Chandra has fixed the accretion rate, and
since the results are insensitive to $\alpha$ and the details of the
magnetic field strength, the model is almost parameter-free.  The only
parameter that has been adjusted is $\delta$, and even this parameter
has a reasonably wide range of allowed values.

As an aside, we note that the argument we made in \S3 (second
paragraph) against the presence of a ``dead'' thin disk in Sgr A$^*$
does not apply to the ADAF model.  Since the ADAF extends all the way
out to the capture radius, there is no supersonically infalling
region, and therefore there is no shock where the infalling gas may
thermalize and radiate the incoming kinetic energy.  Another point is
that the ADAF model does not require the incoming gas to have a high
specific angular momentum.  As discussed in \S4.2, so long as there is
sufficient angular momentum to prevent the gas from falling into the
BH directly, a rotating viscous ADAF will grow to the size of the
capture radius.  Only then can angular momentum be transported out
into the ambient medium.  For the same reason, the CDAF solution
discussed in \S7 again has to extend out to the capture radius.  In
that case, it is not only angular momentum that needs to be
transported to the outside, but also a great deal of energy.

\begin{figure}[b]
\begin{center}
\includegraphics[width=.6\textwidth]{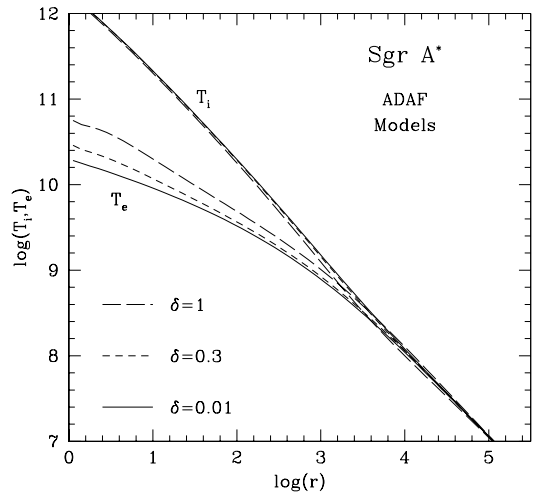}
\end{center}
\caption[]{Variation of the ion and electron temperatures with radius
in the three ADAF models shown in Fig. 4}
\label{sgradaft}
\end{figure}
Figure 5 shows the profiles of the ion and electron temperatures in
the ADAF models discussed above.  These profiles have been calculated
self-consistently (see \cite{Netal98}), with separate energy equations
for the ions and the electrons \cite{NKMK97}, and assuming that the
two specias couple only via Coulomb collisions.  Notice how a
two-temperature structure develops naturally in this model, so that
the electrons do not get any hotter than few$\times 10^{10}$ K.  As
already discussed in \S4, this is the kind of electron temperature
needed to fit the spectral data.  The brightness temperature of Sgr
A$^*$ at mm wavelengths is also measured to be of this order
\cite{Detal01}.

\subsection{The Flare in Sgr A$^*$}
The long-dashed curve in Fig. 4 shows the spectrum of a model with
$\delta=1$, with all the viscous heat assumed to go into the
electrons.  The electrons are hotter in this model than in the other
two models.  Correspondingly, both the synchrotron peak and the
Compton-scattering component are significantly stronger.  The
bremsstrahlung emission is not affected since it is emitted from large
radii, where $\delta$ has very little effect on the temperature
(Fig. 5).

It is interesting that the model with $\delta=1$ fits both the X-ray
flux and the X-ray spectral slope of the flare state of Sgr A$^*$.
Moreover, the spectrum passes through the 2.2 $\mu$m upper limit.  As
discussed in \S2.1, this limit corresponds to a claimed detection of
Sgr A$^*$.  Could it be that Sgr A$^*$ underwent a flare during that
particular observation?

It is certainly a coincidence that the $\delta=1$ model fits the flare
data so well, all the more so since the models are not very reliable
in this limit.  However, the result does show that any minor
perturbation that heats up the electrons at small radii by as little
as a factor of 2 will produce a spectrum similar to that seen in the
flare.  Such a temperature change could be caused by many effects.
Apart from an increase in $\delta$, other possibilities are (i) a
sudden enhancement in the coupling between ions and electrons, (ii) a
major reconnection event, or (iii) a sudden injection of a population
of nonthermal electrons.

\subsection{Does Sgr A$^*$ Have an Event Horizon?}
Even allowing for the fact that the accretion rate in an ADAF model is
smaller than in the Bondi model by a factor $\sim\alpha$, Sgr A$^*$ is
still highly underluminous: $L_{acc}\ll0.1\dot M_{ADAF}c^2$.  Energy
advection is the reason why the radiative efficiency of the accretion
flow is so low.  But advection alone is not enough, because we still
have to decide what happens to the advected energy when it finally
reaches the center.  In the models shown in Fig. 4, it was assumed
that the mass at the center is a BH which swallows the advected
energy.

\begin{figure}[b]
\begin{center}
\includegraphics[width=.6\textwidth]{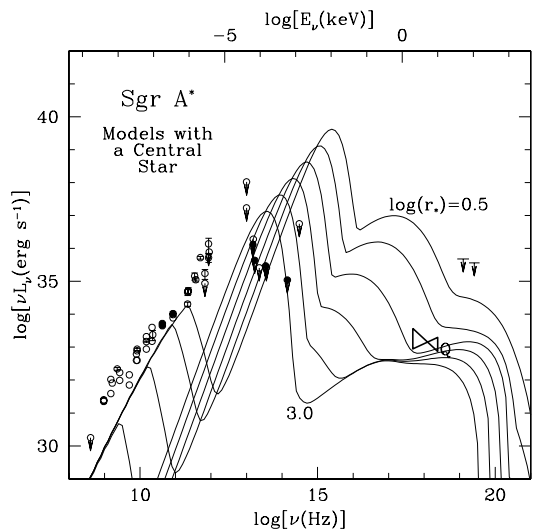}
\end{center}
\caption[]{Spectra of models in which accretion occurs via an ADAF and
the central accreting mass is assumed to have a surface that radiates
the advected energy with a blackbody spectrum.  The curves are labeled
by the radius $r_*$ of the central mass in units of the Schwarzschild
radius.}
\label{sgrstar}
\end{figure}
What would happen if the object were not a BH?  If the object had a
hard surface, then the advected energy would be radiated when the hot
gas hit the surface (because the density would go up and the radiative
efficiency would increase suddenly), and the predicted spectrum would
disagree violently with the data \cite{Netal98},\cite{MQN99}.  Figure
6 shows some results.  The various curves correspond to ADAF models in
which the object in the center is postulated to have a hard surface
with radius equal to $10^{0.5}, 10^{1}, 10^{1.5}, 10^{2}, 10^{2.5},
10^{3}$ Schwarzschild radii.  Notice that all the models are ruled out
by the infrared limits.

We thus conclude that, if the ADAF model is the correct description of
the accretion flow in Sgr A$^*$, then the central object must be a
black hole with an event horizon.  The same is true even if the
accretion proceeds via a Bondi flow.

\section{Other Dim Galactic Nuclei}

\subsection{Nuclei of Giant Ellipticals}
It has long been recognized that the nuclei of nearby giant elliptical
galaxies are unusually dim \cite{FC88}, with $L_{acc}\ll0.1\dot
M_{Bondi}c^2$.  Soon after the successful application of the
two-temperature ADAF model to Sgr A$^*$ \cite{NYM95}, it was proposed
that the same model would also solve the riddle of dim elliptical
nuclei \cite{FR95},\cite{Retal96},\cite{M97}, \cite{DMF97}.

Until recently, $\dot M_{Bondi}$ in the giant ellipticals was
estimated indirectly by modeling the X-ray cooling flow in the central
regions of the galaxy and extrapolating the model to the nucleus.
However, the Chandra X-ray Observatory has improved matters
significantly by providing in a few ellipticals direct images of hot
X-ray emitting gas close to the capture radius of the SMBH.  Good
quality data are presently available for NGC 1399, NGC 4472, NGC 4636
\cite{Letal01} and NGC 6166 \cite{DMetal01}.  From these observations,
reliable estimates of $\dot M_{Bondi}$ have been obtained, and we are
now in a position to state with great confidence that $L_{acc}$ is
indeed very much less than $0.1\dot M_{Bondi}c^2$ in all these
galactic nuclei.

\begin{figure}[b]
\begin{center}
\includegraphics[width=.6\textwidth]{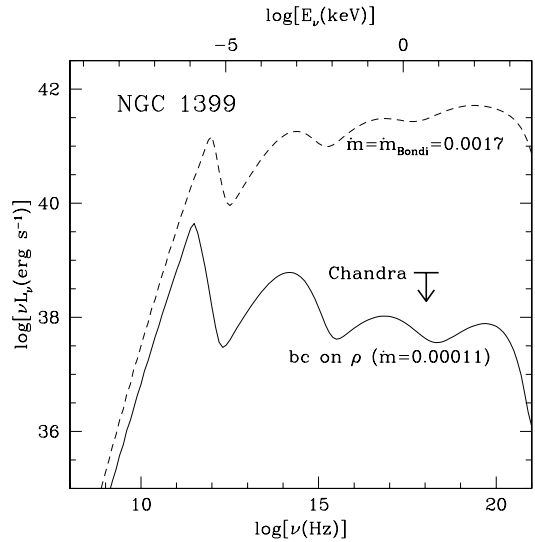}
\end{center}
\caption[]{ADAF models of NGC 1399.  For the dashed curve, $\dot M$
has been set equal to the estimated Bondi accretion rate (0.0017 in
Eddington units).  For the solid line, the ADAF model has been
adjusted so as to match the external density as measured by Chandra
($\dot m=0.00011$).}
\label{NGC1399}
\end{figure}
In fact, $L_{acc}$ is so low and $\dot M_{Bondi}$ is so large that
even a two-temperature ADAF model with $\dot M_{ADAF}=\dot M_{Bondi}$
cannot satisfy the X-ray constraints \cite{Letal01}.  This is shown
for NGC 1399 by the dashed line in Fig. 7.  The predicted spectrum is
far too bright in X-rays compared to the Chandra upper limit.
However, as explained earlier, it is not correct to set $\dot
M_{ADAF}=\dot M_{Bondi}$.  Rather, one must match the density of the
ADAF model to the ambient density of the external medium at $R=R_c$
and solve consistently for $\dot M_{ADAF}$.  When this is done, one
obtains a lower value of $\dot M_{ADAF}$, and correspondingly a
significantly dimmer source.  The solid line in Fig. 7 shows the
spectrum thus obtained for NGC 1399 (using $\alpha=0.1$, but the
results are not very sensitive to this choice).  The model is
consistent with the X-ray data.  Similar results are obtained for
other nuclei for which there is Chandra data.

The matching of the ADAF solution to the external medium was done
somewhat approximately here.  Ideally, one should obtain a consistent
solution of the viscous accretion equations, extending from the
external medium/cooling flow through the capture radius down to the
BH.  The analogous problem for spherical accretion has been solved
\cite{QN00}, but the viscous problem is yet to be analysed.

\subsection{LINERs}
Low-Ionization Nuclear Emission Region (LINER) galaxies, and more
generally low luminosity AGN (LLAGN), have unusual spectral properties
\cite{H99}.  It has been proposed that the observations could be
understood if accretion proceeds via an ADAF in these galactic nuclei
\cite{Letal96},\cite{Qetal99}.  A feature of the proposed models is
that the accretion flow consists of two zones: an outer thin disk that
extends from some large radius down to a transition radius $R_{tr}$
and an inner ADAF that extends from $R_{tr}$ to the BH horizon.

There is some direct evidence for such a geometry of the accretion
flow in NGC 4258 \cite{GNB99}, M81 and NGC 4579 \cite{Qetal99}.  M81
and NGC 4579, in particular, lack the Big Blue Bump that is seen in
AGN.  Instead, they have a red bump, which can be fitted only with a
truncated disk \cite{Qetal99}.  Furthermore, the absence of short time
scale variability \cite{Petal98} and unusually strong radio emission
\cite{UH01} in LLAGN appear to confirm the presence of radiatively
inefficient accretion in these galactic nuclei.

\subsection{Transition From Thin Disk to ADAF}
So far in this article, we have discussed three kinds of accretion
flows.  First, we briefly mentioned bright AGN, with accretion
luminosities of say $L_{acc}\sim 0.1-1 L_{Edd}$.  We said that these
objects have Shakura-Sunyaev thin accretion disks.  Next, at the other
extreme, we discussed Sgr A$^*$, with $L_{acc}\sim10^{-8.5}L_{Edd}$.
We said that this source might have an ADAF extending from the capture
radius at $R\sim10^5R_S$ down to the BH.  Finally, in the previous
subsection, we discussed intermediate cases like M81, with
$L_{acc}\sim10^{-4}-10^{-5}L_{Edd}$.  We argued that the accretion
occurs via a thin disk at large radii ($R>R_{tr}\sim10^2R_S$
\cite{Qetal99}) and vai an ADAF at smaller radii ($R<R_{tr}$).

The pattern suggested by the above facts is very reminiscent of a
paradigm that was developed for BH X-ray binaries
\cite{N96},\cite{EMN97},\cite{Eetal98}.  The key idea is that the
geometry of the accretion flow is determined primarily by the
Eddington-scaled mass accretion rate $\dot m=\dot M/\dot M_{Edd}$.
For $\dot m$ greater than a critical value $\dot
m_{crit}\sim0.01-0.1$, the accretion occurs via a radiatively
efficient thin disk, with perhaps a hot corona.  At these accretion
rates, the ADAF solution is not allowed \cite{NY95b} because the gas
density is too high to permit a radiatively inefficient flow.  Once
$\dot m$ falls below $\dot m_{crit}$, an ADAF is allowed at small
radii and the thin disk develops a small hole in the inner regions
where the cold gas switches to a hot ADAF.  The range of radius over
which an ADAF is allowed increases as $\dot m$ decreases \cite{NY95b},
and correspondingy the hole in the thin disk also becomes larger.  For
an extremely low accretion rate, e.g., $\dot m\sim 10^{-5}$ in Sgr
A$^*$, the hole is so large that the outer thin disk disappears
altogether and the accretion occurs via a pure ADAF over a wide range
of radius.  Such flows are radiatively extremely inefficient.  This
paradigm is in qualitative agreement with a variety of observations,
e.g., \cite{Eetal01}, though undoubtedly there are parameters other
than $\dot m$ that also affect the geometry of the flow.

An important question is: how does the cold gas in a thin disk switch
to a hot quasi-spherical ADAF?  A number of papers have been written
on this ``evaporation'' process
\cite{MM94},\cite{H96},\cite{DT98},\cite{Letal99},\cite{RC00},
\cite{CRZ00},\cite{MLM00},\cite{SD01}, and there is some qualitative
understanding of the relevant physics.  However, there is no reliable
quantitative model yet that can calculate the dependence of the
transition radius $R_{tr}$ on the accretion rate $\dot m$.

\section{Other Possibilities}

\subsection{Bondi Accretion With Magnetic Fields}
Recently, the first three-dimensional numerical simulations of
spherical accretion with magnetic fields were reported \cite{IN01}.
The results turned out to be unexpected.  The density profile of a
magnetized spherical flow was found to differ significantly from the
$R^{-3/2}$ behavior predicted by the Bondi model (\S4), and the mass
accretion rate was found to be much less than the value given in (2).
The reason for the discrepancy is as follows \cite{IN01}.

\begin{figure}[b]
\begin{center}
\includegraphics[width=.6\textwidth]{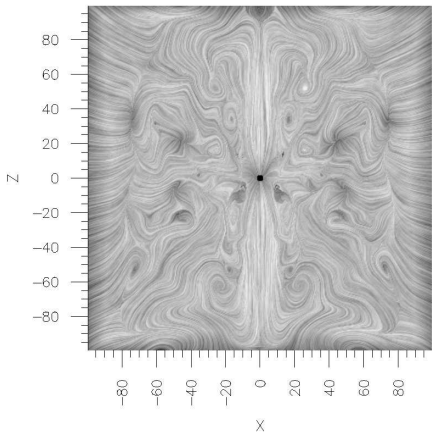}
\end{center}
\caption[]{Velocity streamlines in the meridional plane in a
simulation of a magnetized spherical accretion flow \cite{IN01}.  Note
the obvious turbulent eddies which are due to convection.}
\label{MHDBondi}
\end{figure}
As already noted in \S4, when magnetic fields are frozen into the
inflowing gas, the fields are amplified, causing them to reconnect,
and the energy released thereby heats up the gas.  Since the gas
radiates very little of its energy, the entropy of the gas increases
inward.  The negative radial entropy gradient causes the gas to be
convectively unstable, so violent convection is set up.  Figure 8
shows velocity streamlines in the meridional plane from one of the
numerical simulations.  Notice how turbulent the flow is, and how
different it is from the purely radial streamlines of Bondi's
hydrodynamic model.  In this convective flow, the outward energy flux
due to convection dominates the physics at large radii.  As a result,
the global structure of the flow is strongly affected and the mass
accretion rate onto the BH is much reduced from the Bondi rate.

An analytical model of spherical accretion with convection has been
developed \cite{IN01}.  The model predicts that the density should
vary as $\rho\propto R^{-1/2}$ rather than the Bondi scaling of
$\rho\propto R^{-3/2}$.  The scaling drops out naturally just from the
assumption that the convective energy flux dominates over other
fluxes.  Because of the modified scaling, the accretion rate onto the
BH is much reduced from the Bondi rate (2):
\begin{equation}
\dot M_{MHD}\sim (R_S/R_c)\dot M_{Bondi}.
\end{equation}
The numerical simulations do not have sufficient dynamic range, nor
have they been run long enough, to unequivocally confirm this result.
What is clear from the simulations, however, is that magnetic fields
can modify even such a long-cherished paradigm as the Bondi model.

As discussed in \S4.2, the Bondi model is unlikely to be relevant for
real accretion flows because it ignores the angular momentum of the
gas.  It might therefore appear that the above results are not of
practical importance.  However, as we discuss next, convection causes
similarly large effects even in rotating viscous flows.
(Historically, the rotating flows were studied first.)

\subsection{Convection-dominated Accretion Flow (CDAF)}
An ADAF has two unusual properties that strongly influence the
global structure of the flow.

First, the gas has a positive Bernoulli parameter (sum of potential
energy, kinetic energy and enthalpy), which means that the gas is
technically not bound to the BH and is liable to flow out of the
system \cite{NY94},\cite{NY95a},\cite{NKH97}.  If the outflow is
strong enough, then the amount of mass reaching the BH could be a
great deal less than the mass supplied on the outside
\cite{BB99},\cite{B00} (but see \cite{ALI00},\cite{TD00}).  In such an
``advection-dominated inflow outflow solution'' (ADIOS), the density
profile can be strongly modified: instead of $\rho\propto R^{-3/2}$ as
in a standard ADAF, one could have $\rho\propto R^{-3/2+p}$, with
$0<p<1$ \cite{BB99}.  The limit $p=0$ corresponds to the standard ADAF
model described in \S5.

Second, because viscous dissipation in an ADAF adds heat energy to the
accreting gas, and since there is negligible cooling, the entropy of
the gas increases inward (see the analogous discussion in \S7.1).  As
a result, the gas in an ADAF is convectively unstable
\cite{BM82},\cite{NY94},\cite{NY95a}.  The effect of this convection
has been studied via numerical hydrodynamic simulations
\cite{ICA96},\cite{IA99},\cite{SPB99},\cite{IA00}.  As in the case of
the magnetic Bondi flow, convection alters the density profile to
$\rho\propto R^{-1/2}$ and causes the mass accretion rate onto the BH
to be reduced drastically compared to the standard ADAF model
\cite{SPB99}.  The physics of these ``convection-dominated accretion
flows'' (CDAFs) is fairly well understood
\cite{NIA00},\cite{QG00},\cite{IAN00}.  Convection introduces a strong
flow of energy outward, just as in the magnetized Bondi problem.  In
addition, it also introduces a flow of angular momentum {\it inward}
(i.e. opposite to the direction of viscous transport), as anticipated
in some previous work \cite{RG92},\cite{KNL95},\cite{SB96}.  All of
this leads to novel and quite interesting properties
\cite{NIA00},\cite{QG00}.  Analysis of the global structure of CDAFs
indicates that the gas switches from a convection-dominated state to a
more traditional ADAF-like state for radii less than about $50R_S$
\cite{Aetal01}.

Preliminary work has been done on the effect of magnetic fields on
rotating advection-dominated flows
\cite{Ma99},\cite{H00},\cite{MHM00},\cite{SP01},\cite{MMM01},\cite{HBS01}.
Depending on the initial configuration of the magnetic field, the
dynamical properties of the flow appear to be quite distinct.  For a
predominantly ``vertical'' initial field, there are strong jets
\cite{Ma99}, whereas for a toroidal initial field there are only weak
outflows \cite{MHM00}.  Regardless, convection seems to be present in
both cases, and some studies \cite{MMM01} find quite good agreement
between numerical MHD simulations and analytical work on
hydrodynamical CDAFs.  Others, however, claim that there are large
differences (see especially \cite{B01} where it is argued that
convection in a differentially rotating magnetized medium behaves very
differently from the unmagnetized case).

Work in this area is very important and needs to be pursued
vigorously.  The reason is that pure hydrodynamical simulations are
ultimately limited by the fact that they invoke an artificial
viscosity (parameterised via $\alpha$) to transport angular momentum.
If the shear stress is generated by the magneto-rotational instability
\cite{BH98}, as universally believed, then it is obviously much better
to carry out full MHD simulations so that the magnetic shear stress
and the associated ``viscosity'' are computed self-consistently.

The variability properties of MHD flows have been investigated
\cite{Ketal00},\cite{Metal01}.  It is clear that magnetic flares
associated with reconnection events occurring over a wide range of
scales can give broad-band fluctuations as well as occasional strong
outbursts (as in the Sgr A$^*$ flare).

\subsection{ADIOS/CDAF Models of Sgr A$^*$}
Spectral calculations with the ADIOS model using different values of
the parameter $p$ (which measures the strength of the outflow) are
described in \cite{QN99}.  The results are as follows.  Consider a
sequence of models in which $p$ is varied while keeping the density at
the outer boundary and all other model parameters fixed.  With
increasing $p$, the density and the temperature of the gas near the BH
decrease.  This causes the synchrotron emission and the Compton
emission to drop, without affecting the bremsstrahlung emission (which
comes from the outside).  The good agreement between the ADAF model
and the data in Fig. 4 would then be lost.  However, if the parameter
$p$ as well as the electron heating parameter $\delta$ (\S5.1) are
both increased simultaneously, then one can recover a good fit with
the observations \cite{QN99}.  According to Fig. 4, a standard ADAF
model ($p=0$) with $\delta\sim0.3$ gives a good fit to the data.  By
increasing $p$ and $\delta$ simultaneously, larger values of $p$ are
also likely to give acceptable results.

Although the CDAF model has very different physics than the ADIOS
model, nevertheless, as far as spectral calculations are concerned, it
behaves very much like an ADIOS model with $p=1$.  Spectra
corresponding to CDAF models have been presented in the literature
\cite{BNQ01}, but no results are available specifically for Sgr A$^*$.
It is possible that even with $\delta=1$, i.e. with all the viscous
energy going into the electrons, the spectrum may still be deficient
in the radio/mm band relative to the data.  The X-ray spectrum might
also be a little too soft.  These are not necessarily bad, but they
imply that one would need to include additional components in the
model, or additional physics, to explain the observations.  In the
former category would be jet models such as those described below, and
in the latter category would be models that include nonthermal
electrons.

Although the theoretical motivations behind the ADIOS and CDAF models
are strong, there is as yet no unambiguous observational evidence for
these models in astrophysical sources.  If a claimed detection
\cite{Aetal00} of linear polarization of Sgr A$^*$ in mm waves is
confirmed (but see \cite{BWFB01}), it would strongly suggest that the
density of gas close to the BH is much less than that predicted by the
ADAF model \cite{QG00b},\cite{A00}.  An ADIOS model with a largish
value of $p$ or a CDAF model (which is equivalent to $p=1$) would then
be indicated.  The discovery of radio variability with a possible 106
day cycle \cite{ZBG01} independently suggests the presence of a
turbulent CDAF.  However, none of these indications is particularly
robust at this time.

\subsection{Jets, Other Components}
In the discussion so far, it has been tacitly assumed that the
accretion flow is the source of all the observed radiation.  In bright
AGN, it is known that radio emission is usually associated with
relativistic jets.  Since Sgr A$^*$ is brightest in radio and mm
waves, it is not unreasonable to suppose that some of the observed
radiation at these wavelengths originates in a jet, or some other
component that is external to the accretion flow.  Models of this kind
have been developed by several groups (see \cite{MF01} for a review).

In the context of jet models, we note that a pure accretion model with
an ADAF does a pretty good job of explaining the data on Sgr A$^*$
(Fig. 4).  As explained in \S5.2, the model has almost no free
parameters, since the accretion rate is fixed by the Chandra
observations.  If Sgr A$^*$ does have an ADAF, then there is very
little room for additional emission from a jet.

On the other hand, if the accretion flow corresponds to an ADIOS or a
CDAF, then the accretion flow might produce very little radio or mm
emission, especially if $\delta$ is small.  In this case, almost all
the radio and mm emission could be from something other than the
accretion flow.  Jet models then become attractive (but see
\cite{YMF01}).

Note that most of the quiescent X-ray emission in Sgr A$^*$ is
spatially resolved \cite{Betal01a}.  Therefore, jet models of the
quiescent emission need to ensure that they do not predict too much
emission in X-rays, since the jet emission is expected to be
unresolved.  During the X-ray flare in Sgr A$^*$, all the observed
excess emission was point-like and so this emission could in principle
be entirely from a jet.

Successful disk-plus-jet models have been developed for Sgr A$^*$ in
which synchro-Compton emission from a compact nozzle region of a jet
produces the observed radiation
\cite{FMB93},\cite{F96},\cite{FB99},\cite{FM00}.  The ``disk'' in
these models is unlikely to be a standard thin disk for the reasons
discussed in \S3.  Indeed, recent work suggests that a jet-plus-ADAF
model is able to combine many of the attractive features of both
models \cite{YMF01}.  A jet also provides a convincing explanation for
the flare in Sgr A$^*$ \cite{Maetal01}.  

One interesting result is that the electron energy distribution in Sgr
A$^*$ has to be nearly mono-energetic in order to fit the sharp cutoff
of the spectrum in the infrared \cite{DL94},\cite{BD97}.  This is a
somewhat curious result, since jets in bright AGN almost always have
power-law energy distributions.  If the relativistic electrons in jets
are accelerated via shocks, there must be something very different
about the shock in very low-luminosity systems like Sgr A$^*$.

\section{Summary and Conclusions}
The Chandra X-ray Observatory has eliminated a major uncertainty that
has hampered our understanding of dim galactic nuclei.  Thanks to
Chandra's excellent angular resolution, we now have direct
measurements of the density and temperature of ambient gas close to
the gravitational capture radius of the SMBHs in Sgr A$^*$ and a few
nearby galactic nuclei.  This information allows us to estimate for
these nuclei the Bondi mass accretion rate $\dot M_{Bondi}$, as well
as the accretion rate in an advection-dominated accretion flow $\dot
M_{ADAF}$.  With the uncertainty in $\dot M$ removed, we are now in a
position to answer the question posed in the title of this article.
The answer consists of three parts:

\begin{itemize}
\item
In all the dim galactic nuclei for which Chandra has provided a direct
estimate of $\dot M_{Bondi}$, the accretion rate is found to be well
below the Eddington rate.  This is in contrast to bright AGN which are
believed to accrete at close to the Eddington rate.  Thus, the first,
and obvious, reason why AGN switch off is that the gas supply to the
SMBH is reduced, presumably because most of the gas has been converted
into stars.
\item
Even after allowing for the reduced $\dot M$, the objects studied are
still anomalously dim: $L_{acc}\ll0.1\dot M_{Bondi}c^2$.  Therefore,
we can state with great confidence that the accretion {\it must}
proceed via a radiatively inefficient mode.  A two-temperature ADAF
model fits the available data quite well (Fig. 4), with almost no
adjustable parameters (only $\delta$ needs to be adjusted, and even it
is loosely constrained: \S5.2).  In this model, there is not much room
for additional emission from a jet.  A Bondi model can also be made to
fit the data, provided the accreting gas is taken to be
two-temperature (\S4.2).  However, the neglect of angular momentum of
the gas is a serious weakness of the model.  Both the ADAF and Bondi
models work only if the central object has an event horizon (\S5.4).
Independently of these results, one can state with high confidence
that there is no Shakura-Sunyaev thin disk in Sgr A$^*$ (\S3).
\item
Outflows and convection may be important, in which case the mass
accretion rate onto the BH may be significantly less than the mass
supply on the outside, i.e.  $\dot M_{BH}\ll M_{Bondi},\dot M_{ADAF}$
(see the discussion of ADIOS/CDAF models in \S\S7.1--7.3).  There is
as yet no compelling observational evidence for these models, but
there are strong theoretical reasons for favoring them.  If the
accretion flows in Sgr A$^*$ and other dim nuclei are of the ADIOS or
CDAF type, then the accretion flow may be very dim in radio/mm, and
the observed emission in these bands may come from a relativistic jet
or some other component external to the accretion flow.
\end{itemize}

Thus, it appears that three different effects all conspire to make
nearby galactic nuclei extraodinarily dim: there is less gas
available, the gas accretes via a radiatively inefficient mode, and
(perhaps) less gas reaches the BH than is available for accretion.

\bigskip\noindent \emph{Acknowledgements:} The author thanks Shin
Mineshige and Eliot Quataert for useful comments on the manuscript and
the W.M. Keck Foundation for support as a Keck Visiting Professor at
the Institute for Advanced Study, Princeton.  This research was
supported by NSF grant AST-9820686.

%


\begin{thebibliography}{8.}
\addcontentsline{toc}{section}{References}

\bibitem{K99} J.H. Krolik: \emph{Active Galactic Nuclei} (Princeton
University Press, Princeton 1999)

\bibitem{SS73} N.I. Shakura, S.A. Sunyaev: A\&A, 24, 337 (1973)

\bibitem{NT73} I.D. Novikov, K.S. Thorne: in \emph{Blackholes}, ed. by
C. DeWitt, B. DeWitt (Gordon \& Breach, 1973) p343

\bibitem{P81} J.E. Pringle: ARAA, 19, 137 (1981)

\bibitem{FKR92} J. Frank, A. King, D. Raine: \emph{Accretion Power
in Astrophysics} (Cambridge Univ. Press, Cambridge 1992)

\bibitem{KB99} A. Koratkar, O. Blaes: PASP, 111, 1 (1999)

\bibitem{HM91} F. Haardt, L. Maraschi: ApJ, 380, L51 (1991)

\bibitem{HM93} F. Haardt, L. Maraschi: ApJ, 413, 507 (1993)

\bibitem{Retal98} D. Richstone et al.: Nature, 395A, 14 (1998)

\bibitem{MDZ96} P.G. Mezger, W.J. Duschl, R. Zylka: A\&A Rev., 7, 289
(1996)

\bibitem{Netal98} R. Narayan, R. Mahadevan, J.E. Grindlay, R.G. Popham,
C. Gammie: ApJ, 492, 554 (1998)

\bibitem{MF01} F. Melia, H. Falcke: ARAA, 39, 309 (2001)

\bibitem{Betal01a} F.K. Baganoff et al.: ApJ, in press (2001)
(astro-ph/0102151)

\bibitem{Betal01b} F.K. Baganoff et al.: Nature, 413, 45 (2001)

\bibitem{Getal97} R. Genzel, A. Eckart, T. Ott, F. Eisenhauer: MNRAS,
291, 219 (1997)

\bibitem{Wetal81} M.G. Watson, R. Willingale, J.E. Grindlay, P. Hertz:
ApJ, 250, 142 (1981)

\bibitem{Getal94} A. Goldwurm et al.: Nature, 371, 589 (1994)

\bibitem{PT94} P. Predehl, J. Trumper: A\&A, 290, L29 (1994)

\bibitem{Ketal96} K. Koyama et al.: PASJ, 48, 249 (1996)

\bibitem{MS96} M. Morris, E. Serabyn: ARAA, 34, 645 (1996)

\bibitem{EG97} A. Eckart, R. Genzel: MNRAS, 284, 576 (1997)

\bibitem{Ghetal98} A.M. Ghez, B.L. Klein, M. Morris, E.E. Becklin:
ApJ, 509, 678 (1998)

\bibitem{Getal00} R. Genzel, C. Pichon, A. Eckart, O.E. Gerhard,
T. Ott: MNRAS, 317, 348 (2000)

\bibitem{Ghetal00} A.M. Ghez, M. Morris, E.E. Becklin, A. Tanner,
T. Kremenek: Nature, 407, 349 (2000)

\bibitem{B52} H. Bondi: MNRAS, 112, 195 (1952)

\bibitem{ST83} S.L. Shapiro, S.A. Teukolsky: \emph{Black Holes, White
Dwarfs, and Neutron Stars} (Wiley Interscience, New York, 1983)

\bibitem{CM97} R.F. Coker, F. Melia: ApJ, 488, L149 (1997)

\bibitem{QNR99} E. Quataert, R. Narayan, M.J. Reid: ApJ, 517, 101 (1999)

\bibitem{FM97} H. Falcke, F. Melia: ApJ, 479, 740 (1997)

\bibitem{S71} V.F. Shvartsman: Sov. Astr., 15, 37 (1971)

\bibitem{M75} P. Meszaros: A\&A, 44, 59 (1975)

\bibitem{M92} F. Melia: ApJ, 387, L25 (1992)

\bibitem{M94} F. Melia: ApJ, 426, 577 (1994)

\bibitem{MNY96} R. Mahadevan, R. Narayan, I. Yi: ApJ, 465, 327 (1996)

\bibitem{KM99} V. Kowalenko, F. Melia: MNRAS, 310, 1053

\bibitem{LV99a} A. Lazarian, E.T. Vishniac: ApJ, 511, 193 (1999)

\bibitem{LV99b} A. Lazarian, E.T. Vishniac: ApJ, 517, 700 (1999)

\bibitem{SS75} S.L. Shapiro, E.E. Salpeter: ApJ, 198, 671 (1975)

\bibitem{SLE76} S.L. Shapiro, A.P. Lightman, D.M. Eardley: ApJ, 204,
187 (1976)

\bibitem{BH98} S.A. Balbus, J.F. Hawley: Rev. Mod. Phys., 70, 1 (1998)

\bibitem{MLC00} F. Melia, S. Liu, R. Coker: ApJ, 545, L117 (2000)

\bibitem{P77} J.E. Pringle: MNRAS, 178, 195 (1977)

\bibitem{PN92} R. Popham, R. Narayan: ApJ, 394, 255 (1992)

\bibitem{I77} S. Ichimaru: ApJ, 214, 840 (1977)

\bibitem{Retal82} M.J. Rees, M.C. Begelman, R.D. Blandford, E.S.
Phinney: Nature, 295, 17 (1982)

\bibitem{NY94} R. Narayan, I. Yi: ApJ, 428, L13 (1994)

\bibitem{NY95a} R. Narayan, I. Yi: ApJ, 444, 231 (1995)

\bibitem{NY95b} R. Narayan, I. Yi: ApJ, 452, 710 (1995)

\bibitem{Aetal95} M. Abramowicz, X. Chen, S. Kato, J.P. Lasota,
O. Regev: ApJ, 438, L37 (1995)

\bibitem{Cetal95} X. Chen, M.A. Abramowicz, J.P. Lasota, R. Narayan,
I. Yi: ApJ, 443, L61 (1995)

\bibitem{T01} D. Tsiklauri: New Astron., 6, 487 (2001)

\bibitem{MQ97} R. Mahadevan, E. Quataert: ApJ, 490, 605 (1997)

\bibitem{MNK97} R. Mahadevan, R. Narayan, J. Krolik: ApJ, 486, 268 (1997)

\bibitem{M99} R. Mahadevan: MNRAS, 304, 501 (1999)

\bibitem{OPN00} F. Ozel, D. Psaltis, R. Narayan: ApJ, 541, 234 (2000)

\bibitem{EMN97} A.A. Esin, J.E. McClintock, R. Narayan: ApJ, 489, 865;
500, 523 (1997)

\bibitem{Yetal00} F. Yuan, Q. Peng, J. Lu, J. Wang: ApJ, 537, 236
(2000) ???

\bibitem{P78} T. Piran: ApJ, 221, 652 (1978)

\bibitem{KAC96} S. Kato, M.A. Abramowicz, X. Chen: PASJ, 48, 67 (1996)

\bibitem{Ketal97} S. Kato, T. Yamasaki, M.A. Abramowicz, X. Chen:
PASJ, 49, 221 (1997)

\bibitem{WL96} X. Wu, Q. Li: ApJ, 469, 776 (1996)

\bibitem{W97} X. Wu: MNRAS, 292, 113 (1997) 

\bibitem{NMQ98} R. Narayan, R. Mahadevan, E. Quataert: in \emph{The
Theory of Black Hole Accretion Discs}, ed. by M.A. Abramowicz,
G. Bjornsson, J.E. Pringle (Cambridge Univ. Press, Cambridge, 1998)
p148

\bibitem{Ketal98} S. Kato, J. Fukue, S. Mineshige: \emph{Black Hole
Accretion Disks} (Kyoto Univ. Press, Kyoto, 1998)

\bibitem{Q01} E. Quataert: in \emph{Probing the Physics of Active
Galactic Nuclei by Multiwavelength Monitoring}, ed. by B.M. Peterson,
R.S. Polidan, R.W. Pogge (Astr. Soc. Pacific, San Francisco, 2001) p71

\bibitem{NGM01} R. Narayan, M.R. Garcia, J.E. McClintock: in
\emph{Proc. IX Marcel Grossmann Conference}, in press (2001)
(astro-ph/0107387)

\bibitem{NYM95} R. Narayan, I. Yi, R. Mahadevan: Nature, 374, 623 (1995)

\bibitem{MMK97} T. Manmoto, S. Mineshige, M. Kusunose: ApJ, 489, 791
(1997)

\bibitem{M98} R. Mahadevan: Nature, 394, 651 (1998)

\bibitem{M00} T. Manmoto: ApJ, 534, 734 (2000)

\bibitem{Ph81} E.S. Phinney: in \emph{Plasma Astrophysics}, ed. by
T. Guyenne (ESA SP-161, 1981) p337

\bibitem{BL97} G.S. Bisnovatyi-Kogan, R.V.E Lovelace: ApJ, 486, L43 (1997)

\bibitem{Q98} E. Quataert: ApJ, 500, 978 (1998)

\bibitem{G98} A. Gruzinov: ApJ, 501, 787 (1998)

\bibitem{QG99} E. Quataert, A. Gruzinov: ApJ, 520, 248 (1999)

\bibitem{B99} E. Blackman: MNRAS, 302, 723 (1999)

\bibitem{NKMK97} K.E. Nakamura, M. Kusunose, R. Matsumoto, S. Kato:
PASJ, 49, 503 (1997)

\bibitem{Detal01} S.S. Doeleman et al.: AJ, 121, 2610 (2001)

\bibitem{MQN99} K. Menou, E. Quataert, R. Narayan: in \emph{Black
Holes, Gravitational Radiation and the Universe}, ed. by B.R. Iyer,
B. Bhawal (Kluwer, Dordrecht, 1999) p265

\bibitem{FC88} A.C. Fabian, C.R. Canizares: Nature, 333, 829 (1988)

\bibitem{FR95} A.C. Fabian, M.J. Rees: MNRAS, 277. L5 (1995)

\bibitem{Retal96} C.S. Reynolds, T. di Matteo, A.C. Fabian, U. Hwag,
C.R. Canizares: MNRAS, 283, L111 (1996)

\bibitem{M97} R. Mahadevan: ApJ, 477, 585 (1997)

\bibitem{DMF97} T. di Matteo, A.C. Fabian: MNRAS, 286, 50 (1997)

\bibitem{Letal01} M. Loewenstein, R.F. Mushotzky, L. Agnelini, 
K.A. Arnaud, E. Quataert: ApJ, 555, L21 (2001)

\bibitem{DMetal01} T. di Matteo, R.M. Johnstone, A.C. Fabian,
S.W. Allen: ApJ, 550, L19 (2001)

\bibitem{QN00} E. Quataert, R. Narayan: ApJ, 528, 236 (2000)

\bibitem{H99} L.C. Ho: ApJ, 516, 672 (1999)

\bibitem{Letal96} J.P. Lasota, M.A. Abramowicz, X. Chen, J. Krolik,
R. Narayan, I. Yi: ApJ, 462, 142 (1996)

\bibitem{GNB99} C.F. Gammie, R. Narayan, R. Blandford: ApJ, 516, 177
(1999)

\bibitem{Qetal99} E. Quataert, T. di Matteo, R. Narayan, L.C. Ho:
ApJ, 525, L89 (1999)

\bibitem{Petal98} A. Ptak, T. Yaqoob, R. Mushotzky, P. Serlemitsos,
R. Griffiths: ApJ, 501, L37 (1998)

\bibitem{UH01} J.S. Ulvestad, L.C. Ho: ApJ, 562, L133 (2001)

\bibitem{N96} R. Narayan: ApJ, 461, 136 (1996)

\bibitem{Eetal98} A.A. Esin, R. Narayan, W. Cui, J.E. Grove,
S.N. Zhang: ApJ, 505, 854 (1998)

\bibitem{Eetal01} A.A. Esin, J.E. McClintock, J.J. Drake, M.R. Garcia,
C.A. Haswell, R.I. Hynes, M.P. Muno: ApJ, 555, 483 (2001)

\bibitem{MM94} F. Meyer, E. Meyer-Hofmeister: A\&A, 361, 175 (1994)

\bibitem{H96} F. Honma: PASJ, 48, 77 (1996)

\bibitem{DT98} C.P. Dullemond, R. Turolla: ApJ, 503, 361 (1998)

\bibitem{Letal99} B.F. Liu, W. Yuan, F. Meyer, E. Meyer-Hofmeister,
G.Z. Xie: ApJ, 527, L17 (1999)

\bibitem{RC00} A. Rozanska, B. Czerny: A\&A, 360, 1170 (2000)

\bibitem{CRZ00} B. Czerny, A. Rozanska, P.Y. Zycki: New Astron. Rev., 
44, 439 (2000)

\bibitem{MLM00} F. Meyer, B.F. Liu, E. Meyer-Hofmeister: A\&A, 361,
175 (2000)

\bibitem{SD01} H.C. Spruit, B. Deufel: A\&A, submitted (2001)
(astro-ph/0108497)

\bibitem{IN01} I.V. Igumenshchev, R. Narayan: ApJ, in press (2001)
(astro-ph/0105365)

\bibitem{NKH97} R. Narayan, S. Kato, F. Honma: ApJ, 476, 49 (1997)

\bibitem{BB99} R.D. Blandford, M.C. Begelman: MNRAS, 303, L1 (1999)

\bibitem{B00} T. Beckert: ApJ, 539, 223 (2000)

\bibitem{ALI00} M.A. Abramowicz, J.P. Lasota, I.V. Igumenshchev:
MNRAS, 314, 775 (2000)

\bibitem{TD00} R. Turolla, C.P. Dullemond: ApJ, 531, L49 (2000)

\bibitem{BM82} M.C. Begelman, D.L. Meier: ApJ, 253, 873 (1982)

\bibitem{QN99} E. Quataert, R. Narayan: ApJ, 520, 298 (1999)

\bibitem{ICA96} I.V. Igumenshchev, X. Chen, M.A. Abramowicz: MNRAS,
278, 236 (1996)

\bibitem{IA99} I.V. Igumenshchev, M.A. Abramowicz: MNRAS, 303, 309 (1999)

\bibitem{SPB99} J.M. Stone, J.E. Pringle, M.C. Begelman: MNRAS, 310,
1002 (1999)

\bibitem{IA00} I.V. Igumenshchev, M.A. Abramowicz: ApJ, 537, L27 (2000)

\bibitem{NIA00} R. Narayan, I.V. Igumenshchev, M.A. Abramowicz:
ApJ, 539, 798 (2000)

\bibitem{QG00} E. Quataert, A. Gruzinov: ApJ, 539, 809 (2000)

\bibitem{IAN00} I.V. Igumenshchev, M.A. Abramowicz, R. Narayan: ApJ,
537, L27 (2000)

\bibitem{RG92} D. Ryu, J. Goodman: ApJ, 388, 438 (1992)

\bibitem{KNL95} P. Kumar, R. Narayan, A. Loeb: ApJ, 453, 480 (1995)

\bibitem{SB96} J.M. Stone, S.A. Balbus: ApJ, 464, 364 (1996)

\bibitem{Aetal01} M.A. Abramowicz, I.V. Igumenshchev, E. Quataert,
R. Narayan: ApJ, in press (2001) (astro-ph/0110371)

\bibitem{Ma99} R. Matsumoto: in \emph{Numerical Astrophysics},
ed. by S.M. Miyama, K. Tomisaka, T. Hanawa (Kluwer, Dordrecht, 1999),
p195

\bibitem{H00} J.F. Hawley: ApJ, 528, 462 (2000)

\bibitem{MHM00} M. Machida, M.R. Hayashi, R. Matsumoto: ApJ, 532, L67
(2000)

\bibitem{SP01} J.M. Stone, J.E. Pringle: MNRAS, 322, 461 (2001)

\bibitem{MMM01} M. Machida, R. Matsumoto, S. Mineshige: PASJ, 53, L1 (2001)

\bibitem{HBS01} J.F. Hawley, S.A. Balbus, J.M. Stone: ApJ, 554, L49 (2001)

\bibitem{B01} S.A. Balbus: ApJ, 562, 909 (2001)

\bibitem{Ketal00} T. Kawaguchi, S. Mineshige, M. Machida, R. Matsumoto,
K. Shibata: PASJ, 52, L1 (2000)

\bibitem{Metal01} S. Mineshige, H. Negoro, R. Matsumoto, M. Machida: this
volume (2001)

\bibitem{BNQ01} G.H. Ball, R. Narayan, E. Quataert: ApJ, 552, 221 (2001)

\bibitem{Aetal00} D.K. Aitken et al.: ApJ, 534, L173 (2000)

\bibitem{QG00b} E. Quataert, A. Gruzinov: ApJ, 545, 842 (2000)

\bibitem{A00} E. Agol: ApJ, 538, L121 (2000)

\bibitem{ZBG01} J.H. Zhao, G.C. Bower, W.M. Goss: ApJ, 547, L29 (2001)

\bibitem{BWFB01} G.C. Bower, M.C.H. Wright, H. Falcke, D.C. Backer:
ApJ, 555, L103 (2001)

\bibitem{F01} H. Falcke: in \emph{Reviews in Modern Astronomy 14:
Dynamical Stability and Instabilities in the Universe}, ed. by
R.E. Schielicke (Astronomische Gesellschaft, Hamburg, 2001) p15

\bibitem{YMF01} F. Yuan, S. Markoff, H. Falcke: A\&A, in press (2001)
(astro-ph/0112464)

\bibitem{FMB93} H. Falcke, K. Mannheim, P.L. Biermann: A\&A, 278, 71
(1993)

\bibitem{F96} H. Falcke: in \emph{IAU Symp. 169: Unsolved Problems in
the Milky Way}, ed. by L. Blitz, P.J. Teuben (Kluwer, Dordrecht, 1996)
p163

\bibitem{FB99} H. Falcke, P.L. Biermann: A\&A, 342, 49 (1999)

\bibitem{FM00} H. Falcke, S. Markoff: A\&A, 362, 113 (2000)

\bibitem{Maetal01} S. Markoff, H. Falcke, F. Yuan, P.L. Biermann: A\&A,
379, L13 (2001)

\bibitem{DL94} W.J. Duschl, H. Lesch: A\&A, 286, 431 (1994)

\bibitem{BD97} T. Beckert, W.J. Duschl: A\&A, 328, 95 (1997)


\end{thebibliography}
\end{document}